\documentclass[aps,prd,onecolumn,groupedaddress,showpacs,nofootinbib,amssymb
]{revtex4}
\usepackage[dvips]{graphicx}
\usepackage{amssymb}
\usepackage{amsmath}
\usepackage{graphicx,,color}
\usepackage{amsfonts}
\usepackage{bm}
\usepackage{cancel}
\usepackage{comment}

\allowdisplaybreaks[4]

\begin{document}

\tolerance=5000

\newcommand\be{\begin{equation}}
\newcommand\ee{\end{equation}}
\newcommand\nn{\nonumber \\}
\newcommand\e{\mathrm{e}}

\title{BRS structure of Simple Model of Cosmological Constant and 
Cosmology}

\author{Taisaku Mori$^{1}$\footnote{
E-mail address: mori.taisaku@k.mbox.nagoya-u.ac.jp}, 
Daisuke Nitta$^{1}$\footnote{
E-mail address: 
nitta.daisuke@g.mbox.nagoya-u.ac.jp}
Shin'ichi Nojiri$^{1, 2,}$\footnote{E-mail address:
nojiri@phys.nagoya-u.ac.jp}
}

\affiliation{
$^1$ Department of Physics, Nagoya University, Nagoya
464-8602, Japan \\
$^2$ Kobayashi-Maskawa Institute for the Origin of Particles and
the Universe, Nagoya University, Nagoya 464-8602, Japan}

\begin{abstract}

In arXiv:1601.02203, a simple model has been proposed in order to solve one of the problems 
related with the cosmological constant. The model is 
{induced from}
a topological field theory 
and the model has an infinite numbers of the BRS symmetries. 
The BRS symmetries are, however, spontaneously broken in general. 
In this paper, we investigate the BRS symmetry in more details and show that there is one and 
only one BRS symmetry which is not broken and the unitarity can be guaranteed. 
In the model, the quantum problem of the vacuum energy, which may be identified with the 
cosmological constant, reduces to the classical problem of the initial condition. 
In this paper, we investigate the cosmology given by the model and specify the region 
of the initial conditions which could be consistent with the evolution of the universe. 
We also show that there is a stable solution describing the de Sitter space-time, which may 
explain the accelerating expansion in the current universe. 

\end{abstract}


\maketitle

Recent observations tells that the expansion of the present universe is 
accelerating. 
The energy density generating the accelerating expansion is called as dark energy. 
The simplest model of the dark energy could be a cosmological term with a 
small cosmological constant, $\Lambda^{1/4} \sim 10^{-3}\,\mathrm{eV}$.
The cosmological term can be regarded with the energy density of the 
vacuum but as well-known, the corrections from the matters in the 
quantum field theory to the vacuum energy $\rho_\mathrm{vacuum}$ 
diverges and it is necessary to introduce the cutoff scale 
$\Lambda_\mathrm{cutoff}$, which might be the Planck scale, 
to regularize the divergence. 
Then the obtained value of the vacuum energy 
$\sim \Lambda_\mathrm{cutoff}^4$ is be much larger than 
the observed value  $\left( 10^{-3}\, \mathrm{eV} \right)^4$ 
of the energy density in the universe.  
Even if we impose the supersymmetry in the high energy, 
the vacuum energy by the quantum corrections is evaluated as  
$\sim \Lambda_\mathrm{cutoff}^2 \Lambda_{\cancel{\mathrm{SUSY}}}^2$. 
Here we denote the scale of the supersymmetry breaking by 
$\Lambda_{\cancel{\mathrm{SUSY}}}$. 
Then anyway the vacuum energy coming from the quantum corrections 
could be very large. 
We may use the counter term in order to obtain the observed 
very small vacuum energy $\left(10^{-3}\, \mathrm{eV}\right)^4$ but 
very very fine-tuning is necessary and it looks extremely unnatural. 
About the discussion why the vaccum energy is so small but does not vanish, 
see \cite{Burgess:2013ara} for example. 
Unimodular gravity theories \cite{Anderson:1971pn,
Buchmuller:1988wx,Buchmuller:1988yn, Henneaux:1989zc,Unruh:1988in,
Ng:1990xz,Finkelstein:2000pg,Alvarez:2005iy,Alvarez:2006uu,
Abbassi:2007bq,Ellis:2010uc,Jain:2012cw,
Singh:2012sx,Kluson:2014esa,Padilla:2014yea,Barcelo:2014mua,
Barcelo:2014qva,Burger:2015kie,
Alvarez:2015sba,Jain:2012gc,Jain:2011jc,Cho:2014taa,Basak:2015swx,
Gao:2014nia,Eichhorn:2015bna,
Saltas:2014cta,Nojiri:2015sfd}. were proposed to solve this problem. 
For other secnarios to solve the cosmological constant problems, see  
\cite{Kaloper:2013zca,Kaloper:2014dqa,Kaloper:2015jra, 
Batra:2008cc,Shaw:2010pq,Barrow:2010xt,Carballo-Rubio:2015kaa}
for example. 

In \cite{Nojiri:2016mlb}, motivated by the unimodular gravity theories, 
a new model has been proposed. 
The action of this model is given by, 
\be
\label{CCC7} 
S' = \int d^4 x \sqrt{-g} \left\{ \mathcal{L}_\mathrm{gravity} 
 - \lambda + \partial_\mu \lambda \partial^\mu \varphi  
 - \partial_\mu b \partial^\mu c \right\} 
+ S_\mathrm{matter} \, .
\ee
Here $\lambda$ and $\varphi$ are scalar fields and 
$b$ and $c$ are also scalar fields but they are 
fermionic (Grassmann odd) and later $b$ is identified with the anti-ghost 
and c with ghost. 
The action without the ghost $c$ and anti-ghost $b$ has appeared 
in \cite{Shlaer:2014gna} for other purpose. 
Recently the cosmological perturbation based on the model in (\ref{CCC7}) 
was investigated in \cite{Saitou:2017zyo}. 
In (\ref{CCC7}), we express the action of matters by $S_\mathrm{matter}$ 
and the Lagrangian density of the gravity $\mathcal{L}_\mathrm{gravity}$ 
can be that of arbitrary model. 
We should note that there is not any parameter except the parts coming 
from  $S_\mathrm{matter}$ and $\mathcal{L}_\mathrm{gravity}$ 

We divide the gravity Lagrangian density $\mathcal{L}_\mathrm{gravity}$ 
into the sum of some constant $\Lambda$, which may include the large 
quantum corrections, and other part $\mathcal{L}_\mathrm{gravity}^{(0)}$ 
as $\mathcal{L}_\mathrm{gravity} = \mathcal{L}_\mathrm{gravity}^{(0)} 
 - \Lambda$. 
We also redefine the scalar field $\lambda$ by $\lambda \to \lambda - \Lambda$. 
Then the action (\ref{CCC7}) is rewritten as, 
\be
\label{CCC7R} 
S' = \int d^4 x \sqrt{-g} \left\{ \mathcal{L}_\mathrm{gravity}^{(0)}
 - \lambda + \partial_\mu \lambda \partial^\mu \varphi  
 - \partial_\mu b \partial^\mu c \right\} 
+ S_\mathrm{matter} \, .
\ee
Then the obtained action (\ref{CCC7R}) does not include the constant 
$\Lambda$, which tells that the constant $\Lambda$ does not affect 
the dynamics. 
Although the constant $\Lambda$ may include the large quantum 
corrections from matters to the vacuum energy, the large quantum 
corrections can be tuned to vanish. 

The model in (\ref{CCC7}) includes ghosts \cite{Nojiri:2016mlb}, 
which generates the negative 
norm states in the quantum theory and therefore the model is 
inconsistent but the negative norm states can be excluded by defining 
the physical states by using the BRS symmetry \cite{Becchi:1975nq}. 
In fact, the action is invariant under the infinite numbers of 
the BRS transformation, 
\be
\label{CCC8BR}
\delta \lambda = \delta c = 0\, , \quad 
\delta \varphi = \epsilon c \, , \quad 
\delta b = \epsilon \left( \lambda - \lambda_0 \right)\, .
\ee
Here $\epsilon$ is a fermionic parameter and $\lambda_0$ is a solution 
of the equation, 
\be
\label{lambda0}
0 = \nabla^\mu \partial_\mu \lambda\, ,
\ee
which can be obtained by the variation of the action (\ref{CCC7}) 
with respect to $\varphi$.\footnote{
The existence of the BRS transformation where $\lambda_0$ satisfies Eq.~(\ref{lambda0}) 
was pointed out by R. Saitou.  
} 
If we define the physical states as the states invariant under the BRS 
transformation in (\ref{CCC8BR}), we can consistently exclude 
the negative norm states as in the gauge theory 
\cite{Kugo:1977zq,Kugo:1979gm}. 
By assigning the ghost number $1$ for $c$ and $-1$ for $b$ and $\epsilon$, 
we find that the ghost number is also conserved. 
The four kinds of fields $\lambda$, $\varphi$, $b$, and $c$ can be identified 
with a quartet in Kugo-Ojima's quartet mechanism 
in the gauge theory \cite{Kugo:1977zq,Kugo:1979gm}. 

We should note that the Lagrangian density in the action (\ref{CCC7}), 
\be
\label{SCCP1}
\mathcal{L} =  - \lambda + \partial_\mu \lambda 
\partial^\mu \varphi  - \partial_\mu b \partial^\mu c \, ,
\ee
can be regarded as the Lagrangian density of a topological field theory 
\cite{Witten:1988ze}, where the Lagrangian density is BRS exact, that is, 
given by the BRS transformation of some quantity. 
We may start with the field theory only including the scalar field $\varphi$ 
but the Lagrangian density vanishes $\mathcal{L}_\varphi=0$. 
Because the Lagrangian density vanishes, under any transformation of 
$\varphi$, the action is trivially invariant. 
In this sense, we may regard this theory as a gauge theory. 
We now impose the following gauge condition in order to fix the gauge 
symmetry, 
\be
\label{CCC9}
1 + \nabla_\mu \partial^\mu \varphi = 0\, .
\ee
Then the gauge-fixing Lagrangian \cite{Kugo:1981hm} is given by the BRS 
transformation (\ref{CCC8BR}) of 
$- b \left( 1 + \nabla_\mu \partial^\mu \varphi \right)$.
In fact, we find 
\be
\label{SCCP2R}
\delta \left(- b \left( 1 + \nabla_\mu 
\partial^\mu \varphi \right) \right)
= \epsilon \left( - \left(\lambda - \lambda_0 \right)  
\left( 1 + \nabla_\mu \partial^\mu \varphi \right) 
+ b \nabla_\mu \partial^\mu c \right) 
= \epsilon \left( \mathcal{L} + \lambda_0 
+ \left(\mbox{total derivative terms}\right) 
\right)\, .
\ee
Therefore the Lagrangian density (\ref{SCCP1}) is surely BRS exact up to 
the total derivative terms if $\lambda_0=0$ and we find that the theory in (\ref{SCCP1}) 
{
could be regarded with a topological field theory. 
}
We should note that for the unbroken BRS symmetry, where $\lambda_0 \neq 0$ in general, 
the Lagrangian density (\ref{SCCP1}) is not BRS exact. 
{
In this sense, the Lagrangian density  (\ref{SCCP1}) is not that of the exact topological 
field theory, 
}
which might be a reason 
why the Lagrangian density  (\ref{SCCP1}) gives non-trivial and physically 
relevant contributions. 

We should note that the gauge condition (\ref{CCC9}) does not fix the 
gauge symmetry completely and there remains the residual gauge symmetry. 
In fact, the gauge condition (\ref{CCC9}) 
is invariant under the residual gauge transformation,
\be
\label{rsdlgt}
\varphi \to \varphi + \delta \varphi\, , 
\ee
Here $\delta \varphi$ satisfies the equation 
$\nabla_\mu \partial^\mu \delta \varphi = 0$. 
Then by using the residual gauge symmetry, 
we can choose (restrict to be) the initial condition where $\varphi$ 
is a constant or even zero.\footnote{
The argument comes from the discussions with S. Akagi.}

We should also note that Eq.~(\ref{CCC8BR}) tells that $\lambda$ is nothing 
but the Nakanishi-Lautrup field 
\cite{Nakanishi:1966zz,Nakanishi:1973fu,Lautrup:1967zz}. 
Then by using Eq.~(\ref{CCC8BR}), 
$\lambda - \lambda_0$ is BRS exact, which tells that the vacuum 
expectation value of $\lambda - \lambda_0$ must vanish. 
If the vacuum expectation value of $\lambda - \lambda_0$ does not vanish, 
the BRS symmetry is spontaneously broken 
and we may not be able to consistently impose the physical state condition.  
We should note that there is only one unbroken BRS symmetry in 
the infinite numbers of the BRS symmetry in (\ref{CCC8BR}). 
Because Eq.~(\ref{lambda0}) is the field equation for $\lambda$, 
the real world should be realized by one and only one solution of 
(\ref{lambda0}) for $\lambda$. 
Therefore in the real world, only one $\lambda_0$ is chosen so that 
$\lambda=\lambda_0$ and the 
corresponding BRS symmetry is not broken. 
Therefore by using the unbroken BRS symmetry, we can exclude the negative 
norm state (ghost states) and the unitarity is guaranteed. 
We should also note that $\lambda_0$ can include the classical fluctuation 
as long as $\lambda_0$ satisfies the classical equation (\ref{lambda0}). 
Therefore although the quantum fluctuations are prohibited by the BRS symmetry, 
there could appear the classical fluctuations.

The above arguments tell that the quantum problem of the 
cosmological constant or vacuum energy might be solved. 
There is not, however, any principle to determine the value of $\lambda$ 
or $\Lambda + \lambda$ in the quantum theory. 
The value could be determined by the initial conditions in the classical theory. 
In other words, the quantum problem of the vacuum energy is replaced with the 
classical problem of the initial conditions. 
Then in the following, we investigate the cosmology given by the model (\ref{CCC7}) 
and specify the region of the initial conditions which could be consistent with 
the evolution of the observed universe. 
We may assume the FRW metric with flat spacial part, 
\be
\label{FRW}
ds^2 = - dt^2 + a(t)^2 \sum_{i=1}^3 \left( dx^i \right)^2 \, ,
\ee
and $\lambda$ and $\varphi$ are assumed to only depend on the time 
coordinate $t$. 
In (\ref{FRW}), $a(t)$ is called as the scale factor. 
By the variation of $\lambda$ in the action (\ref{CCC7}), we obtain 
Eq.~(\ref{CCC9}), which has the following form in the FRW metric 
(\ref{FRW}). 
\be
\label{SCCP3}
0= 1 - \left( \frac{d^2 \varphi}{dt^2} 
+ 3 H \frac{d \varphi}{dt} \right) \, .
\ee
Here $H$ is the Hubble rate $H$ defined by $H \equiv \frac{1}{a} \frac{da}{dt}$. 
The general solution of (\ref{SCCP3}) is given by 
\be
\label{SCCP3b}
\varphi (t) = \int^t \frac{dt_1}{a(t_1)^3}\int^{t_1} dt_2 a(t_2)^3
+ \varphi_1 \int^t \frac{dt_1}{a(t_1)^3} +\varphi_2 \, .
\ee
Here, $\varphi_1$ and $\varphi_2$ are some constant. On the other hand, the equation 
given by the variation of $\varphi$ 
is given by (\ref{lambda0}), which has the following form,
\be
\label{SCCP4}
0= \frac{d^2 \lambda}{dt^2} + 3 H \frac{d \lambda}{dt} \, ,
\ee
whose general solution is given by
\be
\label{SCCP4b}
\lambda = \lambda_1 + \lambda_2 \int^t \frac{dt_1}{a(t_1)^3} \, .
\ee
As a gravity theory, we simply consider the Einstein gravity, whose 
Lagrangian density is given by 
\be
\label{SCCP0}
\mathcal{L}_\mathrm{gravity} = \frac{R}{2\kappa^2} - \Lambda\, .
\ee
Here $R$ is the scalar curvature and $\kappa$ is the gravitational 
coupling constant. $\Lambda$ is a cosmological constant but it may include the large 
quantum correction from the matters. 

First by neglecting the contributions from matters, we consider the FRW 
cosmology. 
Then the first and second FRW equations have the following forms:
\begin{align}
\label{SCCP5}
\frac{3}{\kappa^2} H^2 = & \Lambda 
+ \lambda - \frac{d\lambda}{dt} \frac{d\varphi}{dt} \, , \\
\label{SCCP6}
- \frac{1}{\kappa^2} \left( 3 H^2 + 2 \frac{dH}{dt} \right) 
= & - \Lambda - \lambda - \frac{d\lambda}{dt} \frac{d\varphi}{dt} \, .
\end{align}
We can delete $\Lambda$ from Eqs.~(\ref{SCCP5}) and (\ref{SCCP6}) 
and we find,
\be
\label{SCCP7}
\frac{1}{\kappa^2} \frac{dH}{dt} = \frac{d\lambda}{dt} \frac{d\varphi}{dt} \, .
\ee
Then we find that there is a solution, where $\lambda$ is a constant 
$\lambda=\lambda_1$. 
In fact,  $\lambda=\lambda_1$ is a solution of (\ref{SCCP4}) or 
the solution in (\ref{SCCP4b}) with $\lambda_2=0$. 
Then Eq.~(\ref{SCCP7}) tells that $H$ is a constant, $H=H_0$ and therefore 
the space-time is the de Sitter space-time. 
By using (\ref{SCCP5}) or (\ref{SCCP6}), 
we obtain the explicit value of $\lambda=\lambda_1$ as follows,  
\be
\label{SCCP8}
\lambda_1 = - \Lambda + \frac{3 H_0^2}{\kappa^2}\, .
\ee
A solution of Eq.~(\ref{SCCP3}) is given by $\varphi=  \frac{t}{3H_0}$, 
which is s special case in (\ref{SCCP3b}). 
We should note that the value of $H_0$ 
does not depend on the value of the cosmological constant $\Lambda$. 
Because $H_0$ is given by the constant of the integration in (\ref{SCCP7}), 
the value of $H_0$ could be determined by the initial condition or 
something else. 
Then anyway, the value of the cosmological constant $\Lambda$ is irrelevant 
for the cosmology. 
The above result also tells that the problem in the quantum theory for the 
vacuum energy reduces to the initial condition problem in the classical 
heory in our model. 

We now investigate the stability of the solution in (\ref{SCCP8}) 
expressing the de Sitter space-time. 
For this purpose, we consider the pertubation from the solution, 
\be
\label{pert}
H=H_{0}+\delta H\, ,\quad 
\lambda =  - \Lambda + \frac{3 H_0^2}{\kappa^2} 
+ \delta \lambda \, , \quad 
\varphi =  \frac{t}{3H_0} + \delta \varphi \, .
\ee
Then by using (\ref{SCCP3}), (\ref{SCCP4}), and (\ref{SCCP5}), 
we obtain the following equations, respectively, 
\begin{align}
\label{perteq1}
0=& \delta\ddot\varphi+3H_{0}\delta\dot\varphi
 - \frac{1}{3H_0} \delta H \, , \\
\label{perteq2}
0=& \delta\ddot{\lambda}+3H_{0}\delta\dot{\lambda} \, , \\
\label{perteq3}
\frac{6}{\kappa^{2}}H_{0}\delta H =& \delta \lambda
+ \frac{1}{3H_0}\delta \dot{\lambda} \, .
\end{align}
By deleting $\delta H$ from (\ref{perteq1}) and (\ref{perteq3}), we obtain 
\be
\label{perteq4}
0=\delta\ddot\varphi+3H_{0}\delta\dot\varphi
 - \frac{\kappa^{2}}{18H_{0}^2}\left(\delta \lambda 
 + \frac{1}{3H_0} \delta\eta\right) \, .
\ee
Here we have defined a new variable $\delta\eta$ by 
\be
\label{eta}
\delta\eta \equiv \delta\dot\lambda \, ,
\ee
Then we can rewrite (\ref{perteq2}) as follows, 
\be
\label{perteq5}
0=\delta\dot{\eta}+3H_{0}\delta\eta \, .
\ee
By summarizing the equations (\ref{perteq4}), (\ref{eta}), and (\ref{perteq5}), 
we can write the equations in the matrix form, 
\be
\label{perteq6}
\left(
\begin{array}{c}
\delta\dot\lambda\\
\delta\dot{\eta}  \\
\delta\ddot{\varphi} 
\end{array}
\right)
=
A\left(
\begin{array}{c}
\delta\lambda  \\
\delta\eta  \\
\delta\dot{\varphi}  
\end{array}
\right)
\, , \quad 
A\equiv
\left(
\begin{array}{ccc}
0 & 1 & 0   \\
0 & -3H_0 & 0 \\
\frac{\kappa^{2}}{18H_{0}^2} & - \frac{\kappa^{2}}{54H_{0}^3} 
& - 3 H_0 
\end{array}
\right)\ , .
\ee
The eigenvalues of the matrix $A$ is given by 
$-3H_0$ and two $0$'s. 
Because there is not positive eigenvalues, 
the solution is stable or at least quasi-stable. 
Then the solution (\ref{SCCP8}) describing the de Sitter space-time 
might correspond to the accelerating expansion in the current universe. 


We now investigate what could be the initial 
condition corresponding to the value of the vacuum enegy in the present universe. 
After the inflation, the universe passed through the radiation-dominated era and 
the matter-dominated era, and entered into the dark energy-dominated era. 
In the radiation-dominated era and the matter-dominated era, the contributions from 
$\lambda$ and $\varphi$ can be neglected and these scalar fields are expected to 
evolve by following (\ref{SCCP3b}) and (\ref{SCCP4b}). 
In the future of the dark energy-dominated era, the universe is expected to be 
described by the asymptotically de Sitter space-time in (\ref{SCCP8}). 

In the radiation-dominated era, the scale factor is given by 
\be
\label{arad}
a(t)=a_\mathrm{rad}t^{1/2}\, , 
\ee
in the matter-dominated era, 
\be
\label{amat}
a(t)=a_\mathrm{mat}t^{2/3} \, ,
\ee
and the dark energy-dominated era, 
\be
\label{ade}
a(t)=a_{\Lambda} \e^{H_0 \sqrt{\Omega_{\Lambda}}t}\, ,.
\ee
Here $a_\mathrm{rad}$, $a_\mathrm{rad}$, and $a_{\Lambda}$ are constants depending on 
the energy density of the radiation, the matter density,  and the dark energy density, 
respectively. 
We express the value of the Hubble rate $H$ in the current universe by $H_0$ and 
the dark energy density parameter by $\Omega_{\Lambda}$. 

Then by using (\ref{SCCP3b}) and (\ref{SCCP4b}), the scalar fields $\lambda(t)$ 
and $\varphi(t)$ in the radiation-dominated era are given by 
\be
\label{radlambdaphi}
\varphi(t) = \varphi_{\mathrm{rad}}(t) \equiv \varphi_{\mathrm{rad}\, 2}
 -\frac{2\varphi_{\mathrm{rad}\, 1}}{a_{\mathrm{rad}}^{3}}t^{-1/2} +\frac{1}{5}t^{2} \, ,
\quad 
\lambda(t) = \lambda_{\mathrm{rad}}(t) \equiv \lambda_{\mathrm{rad}\, 1} 
 -\frac{2\lambda_{\mathrm{rad}\, 2}}{a_{\mathrm{rad}}^{3}}t^{-1/2} \, .
\ee
On the other hand, in the matter-dominated era and the dark energy-dominated 
era, the scalar fields are given by 
\begin{align}
\label{matlambdaphi}
\varphi(t)=& \varphi_{\mathrm{mat}}(t) \equiv \varphi_{\mathrm{mat}\, 2}
 -\frac{\varphi_{\mathrm{mat}\, 1}}{a_{\mathrm{mat}}^{3}}t^{-1} +\frac{1}{6}t^{2} \, , \quad 
\lambda(t) = \lambda_{\mathrm{mat}}(t) \equiv \lambda_{\mathrm{mat}\, 1}
 -\frac{\lambda_{\mathrm{mat}\, 2}}{a_{\mathrm{mat}}^{3}}t^{-1} \, , \\
\label{DElambdaphi}
\varphi(t)=& \varphi_{\Lambda}(t) \equiv 
\varphi_{\Lambda\, 2} -\frac{\varphi_{\Lambda\, 1}}{3H_{0}
\sqrt{\Omega_{\Lambda}}a_{\Lambda}^3} \e^{ -3H_{0}\sqrt{\Omega_{\Lambda}}t} 
 + \frac{t}{3H_0\sqrt{\Omega_{\Lambda}}}\, , \quad 
\lambda(t) = \lambda_{\Lambda}(t) 
\equiv \lambda_{\Lambda\, 1}
 -\frac{\lambda_{\Lambda\, 2}}{3H_{0}
\sqrt{\Omega_{\Lambda}}a_{\Lambda}^3} \e^{ -3H_{0}\sqrt{\Omega_{\Lambda}}t} \, .
\end{align}
Here $\varphi_{\mathrm{rad}\,1}$, $\varphi_{\mathrm{rad}\,2}$, $\lambda_{\mathrm{rad}\,1}$, 
$\lambda_{\mathrm{rad}\,2}$, $\varphi_{\mathrm{mat}\,1}$, $\varphi_{\mathrm{mat}\, 2}$,
$\lambda_{\mathrm{mat}\,1}$, $\lambda_{\mathrm{mat}\,2}$, $\varphi_{\Lambda\,1}$, 
$\varphi_{\Lambda\, 2}$, $\lambda_{\Lambda\,1}$, and $\lambda_{\Lambda\,2}$ are constants. 

We now use appoximations where the radiation-dominated era transited to the 
matter-dominated era at the time $t=t_1$ and the matter-dominated era to the 
dark-energy dominated era at $t=t_2$. 
We connect the solutions in (\ref{radlambdaphi}), (\ref{matlambdaphi}), and 
(\ref{DElambdaphi}) by imposing the continuities of the values of $\varphi$, $\lambda$, 
$\dot\varphi$, and $\dot\lambda$ at the transit points. 
Then at the point $t=t_1$, we require  
\be
\label{att1}
\varphi_{\mathrm{rad}\, 2}
 -\frac{2\varphi_{\mathrm{rad}\, 1}}{a_{\mathrm{rad}}^{3}}t_1^{-1/2} +\frac{1}{5}t_1^{2}
= \varphi_{\mathrm{mat}\, 2}
 -\frac{\varphi_{\mathrm{mat}\, 1}}{a_{\mathrm{mat}}^{3}}t_1^{-1} +\frac{1}{6}t_1^{2}
 \, , \quad 
\lambda_{\mathrm{rad}\, 1} -\frac{2\lambda_{\mathrm{rad}\, 2}}{a_{\mathrm{rad}}^{3}}t_1^{-1/2} 
= \lambda_{\mathrm{mat}\, 1} -\frac{\lambda_{\mathrm{mat}\, 2}}{a_{\mathrm{mat}}^{3}}t_1^{-1} \, .
\ee
and 
\be
\label{att1der}
\frac{\varphi_{\mathrm{rad}\, 1}}{a_{\mathrm{rad}}^{3}}t_1^{-3/2} +\frac{2}{5}t_1
= \frac{\varphi_{\mathrm{mat}\, 1}}{a_{\mathrm{mat}}^{3}}t_1^{-2} +\frac{1}{3}t_1\, , \quad 
\frac{\lambda_{\mathrm{rad}\, 2}}{a_{\mathrm{rad}}^{3}}t_1^{-3/2} 
= \frac{\lambda_{\mathrm{mat}\, 2}}{a_{\mathrm{mat}}^{3}}t_1^{-2} \, .
\ee
Then we find
\begin{align}
\label{phiradmat}
& \varphi_{\mathrm{mat}\,1} 
= \left( \frac{a_{\mathrm{mat}}}{a_{\mathrm{rad}}}\right)^3t_1^{1/2}\varphi_{\mathrm{rad}\, 1 }
 +\frac{1}{15}a_{\mathrm{mat}}^3 t_1^3 \, , \quad 
\varphi_{\mathrm{mat}\, 2} = \varphi_{\mathrm{rad}\, 2} 
- \frac{t_1^{-1/2}\varphi_{\mathrm{rad}\, 1}}{a_{\mathrm{rad}}^3}
+ \frac{1}{10}t_1^{2} \, , \nn
& \lambda_{\mathrm{mat}\, 2}
= \left( \frac{a_{\mathrm{mat}}}{a_{\mathrm{rad}}} \right)^3 t_1^{1/2}\lambda_{\mathrm{rad}\, 2} \, , \quad 
\lambda_{\mathrm{mat}\, 1} 
= \lambda_{\mathrm{rad}\, 1} -\frac{\lambda_{\mathrm{rad}\, 2}}{a_{\mathrm{rad}}^{3}}t_1^{-1/2} \, .
\end{align}
On the other hand, at the point $t=t_2$, we require 
\begin{align}
\label{att2}
\varphi_{\mathrm{mat}\, 2}
 -\frac{\varphi_{\mathrm{mat}\, 1}}{a_{\mathrm{mat}}^{3}}t_2^{-1} +\frac{1}{6}t_2^{2}
= & \varphi_{\Lambda\, 2} -\frac{\varphi_{\Lambda\, 1}}{3H_{0}
\sqrt{\Omega_{\Lambda}}a_{\Lambda}^3} \e^{ -3H_{0}\sqrt{\Omega_{\Lambda}}t_2} 
 + \frac{t_2}{3H_0\sqrt{\Omega_{\Lambda}}} \, , \nn
\lambda_{\mathrm{mat}\, 1}  -\frac{\lambda_{\mathrm{mat}\, 2}}{a_{\mathrm{mat}}^{3}}t_2^{-1} 
= & \lambda_{\Lambda\, 1}-\frac{\lambda_{\Lambda\, 2}}{3H_0
\sqrt{\Omega_{\Lambda}}a_{\Lambda}^3} \e^{ -3H_0\sqrt{\Omega_{\Lambda}}t_2} \, , 
\end{align}
and 
\be
\label{att2der}
\frac{\varphi_{\mathrm{mat}\, 1}}{a_{\mathrm{mat}}^{3}}t_2^{-2} +\frac{1}{3}t_2
= \frac{\varphi_{\Lambda\, 1}}{a_{\Lambda}^3} \e^{ -3H_{0}\sqrt{\Omega_{\Lambda}}t_2} 
 + \frac{1}{3H_0\sqrt{\Omega_{\Lambda}}} \, , \quad 
\frac{\lambda_{\mathrm{mat}\, 2}}{a_{\mathrm{mat}}^{3}}t_2^{-2} 
= \frac{\lambda_{\Lambda\, 2}}{a_{\Lambda}^3} \e^{ -3H_0 \sqrt{\Omega_{\Lambda}}t_2} \, , 
\ee
and we obtain
\begin{align}
\label{phimatL}
\varphi_{\Lambda\, 1} =& \frac{a_{\Lambda}^3}{a_{\mathrm{mat}}^{3}}t_2^{-2} 
\e^{3H_{0}\sqrt{\Omega_{\Lambda}}t_2}\varphi_{\mathrm{mat}\,1}
- \frac{a_{\Lambda}^3 \e^{3H_{0}\sqrt{\Omega_{\Lambda}}t_2}}
{3H_0\sqrt{\Omega_{\Lambda}}}
 +\frac{1}{3}t_2 a_{\Lambda}^3 \e^{3H_{0}\sqrt{\Omega_{\Lambda}}t_2}
\, , \nn
\varphi_{\Lambda\, 2}  =& \varphi_{\mathrm{mat}\,2}
 - \left( 1 - \frac{1}{3H_0 t_2 \sqrt{\Omega_{\Lambda}}} \right)
\frac{\varphi_{\mathrm{mat}\, 1}}{t_2 a_{\mathrm{mat}}^{3}} 
+ \frac{t_2}{9H_0\sqrt{\Omega_{\Lambda}}}+\frac{1}{6}t_2^2
- \frac{1}{9H_{0}^{2}\Omega_{\Lambda}}\, , \nn
\lambda_{\Lambda\, 2} =& \left(\frac{a_{\Lambda}}{a_{\mathrm{mat}}} \right)^3 t_2^{-2}
\e^{ 3H_0\sqrt{\Omega_{\Lambda}}t_2} \lambda_{\mathrm{mat}\, 2} \, , \quad 
\lambda_{\Lambda\, 1} = \lambda_{\mathrm{mat}\, 1} 
 - \left( 1 - \frac{1}{3H_0 t_2 \sqrt{\Omega_{\Lambda}}} \right)
\frac{\lambda_{\mathrm{mat}\, 2}}{t_2 a_{\mathrm{mat}}^{3}}\, .
\end{align}
By combining the above equations, we find 
\begin{align}
\label{relations}
\lambda_{0}+\Lambda=&\frac{3H^{2}_{c}}{\kappa^{2}}
=\Lambda +\lambda_{\Lambda\, 1} 
 -\frac{\lambda_{\Lambda\, 2}}{3H_0\sqrt{\Omega_{\Lambda}}
a_{\Lambda}^3} \e^{ -3H_0\sqrt{\Omega_{\Lambda}}t_2} \, , \nn
\lambda_{\Lambda\,1}=&\lambda_{\mathrm{rad}1}
 -\frac{\lambda_{\mathrm{rad}2}}{a_{\mathrm{rad}}^{3}}t_{1}^{-1/2}
\left[1+t_{1}t_{2}^{-1}\left(1-\frac{t_{2}^{-1}}{3H_{0}
\sqrt{\Omega_{\Lambda}}}\right)\right]\, , \nn
\lambda_{\Lambda\,2}=& 
\lambda_{\mathrm{rad}\,2}\left(\frac{a_{\Lambda}}{a_{\mathrm{rad}}}\right)^{3}
\e^{3H_{0}\sqrt{\Omega_{\Lambda}}t_{2}}t_{2}^{-2}t_{1}^{1/2} \, ,\nn
\varphi_{\Lambda\,1} =& 
\varphi_{\mathrm{rad}\,1}\left(\frac{a_{\Lambda}}{a_{\mathrm{rad}}}\right)^{3}
t_{2}^{-2}t_{1}^{1/2} \e^{3H_{0}\sqrt{\Omega_{\Lambda}}t_{2}} 
 + \frac{1}{15} a_{\Lambda}^3 t_1^3 t_2^{-2} \e^{3H_{0}\sqrt{\Omega_{\Lambda}}t_2}
- \frac{a_{\Lambda}^3 \e^{3H_{0}\sqrt{\Omega_{\Lambda}}t_2}}
{3H_0\sqrt{\Omega_{\Lambda}}}
 + \frac{1}{3}t_2 a_{\Lambda}^3 \e^{3H_{0}\sqrt{\Omega_{\Lambda}}t_2}
\, , \nn
\varphi_{\Lambda\, 2}  =& \varphi_{\mathrm{rad}\,2}
 - \left\{ \frac{t_1^{-1/2}}{a_\mathrm{rad}^3} 
 + \left( 1 - \frac{1}{3H_0 t_2 \sqrt{\Omega_{\Lambda}}} \right)
\frac{t_1^{1/2}}{t_2 a_{\mathrm{rad}}^{3}} \right\} \varphi_{\mathrm{rad}\, 1}
+ \frac{t_2}{9H_0\sqrt{\Omega_{\Lambda}}} 
- \frac{1}{9H_0^{2}\Omega_{\Lambda}}\nn
&~ - \frac{1}{15} \left( 1 - \frac{1}{3H_0 t_2 \sqrt{\Omega_{\Lambda}}} \right)
\frac{t_1^3}{t_2} 
+ \frac{1}{10}t_1^{2} +\frac{1}{6}t_2^2 \, .
\end{align}
Now we consider the constraints on the scalar fields coming from the 
observations. 
For the purpose, we use the values of the cosmological parameters in \cite{Ade:2015xua}. 
\begin{itemize}
\item The scale factor and the cosmological time when the density of the radiation was 
equal to the density of matter: \\ 
$a_{\mathrm{rm}}=2.8\times10^{-4}$, 
$t_{1}=4.7\times10^{4}\, \mathrm{yr} 
\sim1.5\times10^{12}\, \mathrm{s}=2.3\times10^{27}[\mathrm{eV}^{-1}]$. 
\item The scale factor and the cosmological time when the density of the matter was 
equal to the density of dark energy: \\
$a_{\mathrm{m}\Lambda}=0.75$, 
$t_{2}=9.8\times10^{9}\, \mathrm{yr} \sim3.1\times10^{17}\, \mathrm{s}
=4.7\times10^{32}[\mathrm{eV}^{-1}]$.
\item The cosmological time when the mradiation-dominated era began: \\
$t_{3}=10^{-32}\, \mathrm{s} =10^{-17}[\mathrm{eV}^{-1}]$. 
\item The scale factor and the cosmological time in the current universe: \\
$a_{0}=1$,  $t_{0}=13.5\times10^{9}\, \mathrm{yr} \sim4.3\times10^{17}\, \mathrm{s}
=6.5\times10^{34}[\mathrm{eV}^{-1}]$. 
\item  The Hubble constant in the current universe: \\
$H_{0}=70\, \mathrm{km} \mathrm{s}^{-1}\mathrm{Mpc}^{-1}\sim 2.2\times10^{-18}
\, \mathrm{s}^{-1}=1.5\times10^{-33}[\mathrm{eV}]$. 
\item The density parameters of the radiation, the matter, and the dark energy: \\
$\Omega_{\mathrm{r}}=8.4\times 10^{-5}$, $\Omega_{\mathrm{m}}=0.30$, 
$\Omega_{\Lambda}\sim0.70$. 
\end{itemize}
Then we obtain,
\begin{itemize}
\item $a_{\mathrm{rad}}\sim\left(2H_{0}\sqrt{\Omega_{\mathrm{r}}}\right)^{1/2}
=2.0\times10^{-10}\, \mathrm{s}^{-1/2}=5.1\times10^{-18}[\mathrm{eV}^{1/2}]$
\item $a_{\mathrm{mat}}\sim\left(\frac{3}{2}H_{0}\sqrt{\Omega_{\mathrm{m}}}\right)^{2/3}
\sim5.7\times10^{-13}\,\mathrm{s}^{-2/3}=4.3\times10^{-24}~[\mathrm{eV}^{2/3}]$
\item The critical density: $\rho_{0}=\frac{3H_{0}^{2}}{8\pi G}=5\times10^{-24}\, 
\mathrm{kgm}^{-3}=4.2\times10^{-11}~[\mathrm{eV}^{4}]$. 
\item Newton's gravitational constant: $G \sim 6.6\times10^{-11}\, \mathrm{m}^{3}
\mathrm{kg}^{-1} \mathrm{s}^{-2} =6.7\times10^{-57}~[\mathrm{eV}^{-2}]$. 
\end{itemize}


First constraints could be obtained by requiring $\Lambda+\lambda$ should become a 
constant corresponding to the cosmological constant 
$\Lambda_0 \sim 10^{-11}\, [\mathrm{eV}^4]$, 
\be
\label{desitterconstraint}
\Lambda_{0} \sim \Lambda+\lambda_{\Lambda\, 1}\gg\left|\frac{\lambda_{\Lambda\, 2}}{3H_{0}\sqrt{\Omega_{\Lambda}}a_{\Lambda}^3} \e^{ -3H_{0}\sqrt{\Omega_{\Lambda}}t_{0}}\right|\, ,
\ee
By using (\ref{relations}), we can rewrite the constraints in (\ref{desitterconstraint}) as follows, 
\begin{align}
\label{desitterconstraint1}
10^{-11}\, [\mathrm{eV}^4] \sim& \Lambda 
+ \lambda_{\mathrm{rad}1}-\frac{\lambda_{\mathrm{rad}2}}{a_{\mathrm{rad}}^{3}}t_{1}^{-1/2}
\left[1+t_{1}t_{2}^{-1}\left(1-\frac{t_{2}^{-1}}{3H_{0}
\sqrt{\Omega_{\Lambda}}}\right)\right]
\sim \Lambda+\lambda_{\mathrm{rad}1}-\lambda_{\mathrm{rad}2}
\times \left( 3.1\times10^{65}\, [{\mathrm{eV}}^{\frac{1}{2}}] \right) \, , \\
\label{desitterconstraint2}
10^{-11}\, [\mathrm{eV}^4] \gg & \left|\lambda_{\mathrm{rad}2} \right|\times 
\left(1.8\times 10^{36}\, [{\mathrm{eV}}^{\frac{1}{2}}] \right) \, .
\end{align}
Then Eq.~(\ref{desitterconstraint2}) gives the following constraint,
\be
\label{desitterconstraint3}
\left|\lambda_{\mathrm{rad}2} \right| \ll 10^{-47}\, [{\mathrm{eV}}^5]\, .
\ee
Next constriant requires that the matter should be surely dominant compared 
with the contributions 
from $\lambda$ and $\varphi$ in the matter-dominated era $t_1\ll t \ll t_2$, 
\begin{align}
\label{matterdom}
& \Lambda 
+ \lambda_{\mathrm{rad}\, 1}
 -\frac{\lambda_{\mathrm{rad}\, 2}}{a_{\mathrm{rad}}^{3}}t_1^{-1/2} \left(1+t_{1}t^{-1}\right)\nn
&-\frac{\lambda_{\mathrm{rad}\, 2}}{a_{\mathrm{rad}}^{3}}t_1^{1/2}
\left\{ \left(\frac{t_{1}^{1/2}}{a_{\mathrm{rad}}^{3}}\varphi_{\mathrm{rad}\, 1 }
+\frac{1}{15}t_{1}^{3}\right)t^{-4}
+\frac{1}{3}t^{-1}
 \right\}\ll\rho=\Omega_{\mathrm{m}}\rho_0 a_\mathrm{mat}^{-3} t^{-2}  \, .
\end{align}
We should require that the radiation should be dominant in the radiation-dominated era 
$t\ll t_1$, 
\be
\label{radiationdom}
\Lambda 
+ \lambda_{\mathrm{rad}\, 1} 
 - \frac{\lambda_{\mathrm{rad}\, 2}}{a_\mathrm{rad}^3}   
\left( \frac{\varphi_{\mathrm{rad}\, 1}}{a_\mathrm{rad}^3} t^{-3}
 + \frac{12}{5} t^{-1/2}\right)
\ll\rho=\Omega_{\mathrm{r}}\rho_0 a_\mathrm{rad}^{-4} t^{-2} \, .
\ee
It is not so straightforward to solve the constriants (\ref{matterdom}) and (\ref{radiationdom}) 
in general. 
We may, however, evaluate the constraints as follows. 
When the matter-dominated era transitted to the dark energy-dominated era at $t=t_2$, 
the l.h.s. is almost equal to the r.h.s. by the definition of the transition. 
Each of the terms, except the first constant terms, in the l.h.s. becomes larger when 
$t\to t_1$ and the most dominant term is $t^{-4}$ term. 
Then we may have the following constraint, 
\be
\label{matterdom2}
\left| \lambda_{\mathrm{rad}\, 2} \left( 
\varphi_{\mathrm{rad}\, 1 }
 +\frac{2}{5}a_{\mathrm{rad}}^3 t_1^{5/2}\right) \right| 
\ll \Omega_{\mathrm{m}}\rho_0 
\frac{a_{\mathrm{rad}}^6}{a_{\mathrm{mat}}^3} t_1 \, ,
\ee
that is 
\be
\label{matterdom3}
\left| \lambda_{\mathrm{rad}\, 2} \left( 
\varphi_{\mathrm{rad}\, 1 } 
  + 1.4\times 10^{16}\, \left[ \mathrm{eV}^{-1} \right] \right) \right| 
\ll 10^{-23}\, \left[ \mathrm{eV}^4 \right]\, .
\ee
At the begining of the radiation-dominated era $t=t_3$, $t^{-3}$ term dominates in the 
l.h.s. of  Eq.~(\ref{radiationdom}) and we obtain the following constraint, 
\be
\label{radiationdom2}
\left| \lambda_{\mathrm{rad}\, 2} \varphi_{\mathrm{rad}\, 1} \right| 
\ll \Omega_{\mathrm{r}}\rho_0 a_\mathrm{rad}^2 t_3 \, .
\ee
that is 
\be
\label{radiationdom2}
\left| \lambda_{\mathrm{rad}\, 2} \varphi_{\mathrm{rad}\, 1} \right| 
\ll 10^{-62}\, \left[ \mathrm{eV}^4 \right]\, .
\ee
We may summarize the obtained constraints, 
\begin{align}
\label{constraints}
& \Lambda+\lambda_{\mathrm{rad}1}-\lambda_{\mathrm{rad}2} 
\times \left( 3.1\times10^{65}\, [{\mathrm{eV}}^{-1}] \right)
\sim 10^{-11}\, [\mathrm{eV}^4] \, , \quad 
\left|\lambda_{\mathrm{rad}2} \right| \ll 10^{-47}\, [{\mathrm{eV}}^5]\, , \nn
& \left| \lambda_{\mathrm{rad}\, 2} \left( 
\varphi_{\mathrm{rad}\, 1 } 
+ 1.4\times 10^{16}\, \left[ \mathrm{eV}^{-1} \right] \right) \right| 
\ll 10^{-23}\, \left[ \mathrm{eV}^4 \right]\, , \quad 
\left| \lambda_{\mathrm{rad}\, 2} \varphi_{\mathrm{rad}\, 1} \right| 
\ll 10^{-62}\, \left[ \mathrm{eV}^4 \right]\, .
\end{align}
The first constraint in (\ref{constraints}) or (\ref{desitterconstraint1}) 
seems to tell that we need the fine tuning for the initial conditions.

We now consider more about the initial condition for $\lambda$. 
By choosing $t$ as a present time, $\lambda$ could be expressed as 
\be
\label{lambdaC}
\lambda \sim 10^{-11}\, [\mathrm{eV}^4] \sim 
\lambda_{\mathrm{rad}\, 1} - \frac{\lambda_{\mathrm{rad}\, 2}}{6.4\times 10^{-39} 
[\mathrm{eV}]}\, .
\ee
This may tell
\be
\label{lambdaC2}
\lambda_{\mathrm{rad}\, 1} \sim \left( 10^{-3} \, [\mathrm{eV}] \right)^4 \, , \quad 
\lambda_{\mathrm{rad}\, 2} \sim \left( 10^{-10} \, [\mathrm{eV}] \right)^5 \, .
\ee
Then the obtained value seems to be very small. 
If we assume $\lambda_{\mathrm{rad}\, 1}=0$, which might be unnatural, then by using 
(\ref{radlambdaphi}), we find the value of $\lambda$ at the beginning of the radiation-dominated 
era $t\sim t_3$, 
\be
\label{lambdaC2}
\lambda = \lambda_{\mathrm{rad}}(t_3) \sim \left( 0.1\,[\mathrm{keV}] \right)^4 \, .
\ee
The obtained value might be a little bit more reasonable. 
Then even if $\lambda \sim 10^{-12}\, [\mathrm{eV}^4]$ in the present universe, 
$\lambda \sim \left[\left( 0.1\,[\mathrm{keV}] \right)^4\right]$ at the beginning 
of the radiation-dominated era. 
The converse is not true because $\lambda_{\mathrm{rad}\, 1}\neq 0$ in general: 
If we only require $\lambda \sim \left( 0.1\,[\mathrm{keV}] \right)^4$ at the beginning of 
the radiation-dominated 
era, we may find 
$\lambda \sim \left( 0.1\,[\mathrm{keV}] \right)^4\gg 10^{-12}\, [\mathrm{eV}^4]$
even in the present universe. 

\begin{figure}[h]
\centering
\rotatebox{-90}{\includegraphics[width=15pc]{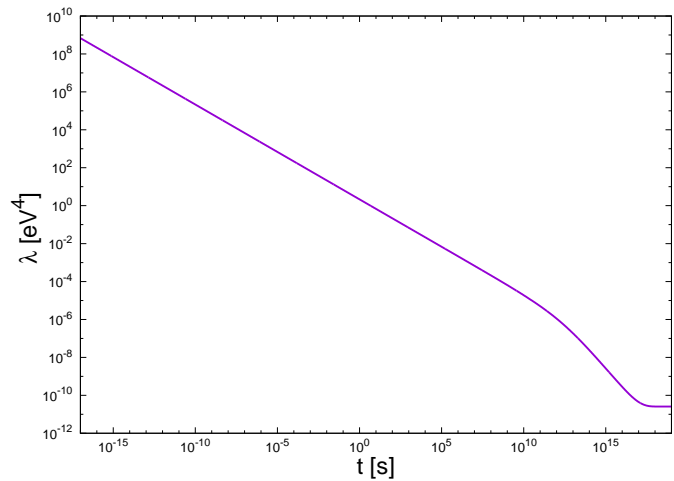}}
\caption{The development of $\lambda$. 
}\label{lambda}
\end{figure}

\begin{figure}[h]
\centering
\rotatebox{-90}{\includegraphics[width=15pc]{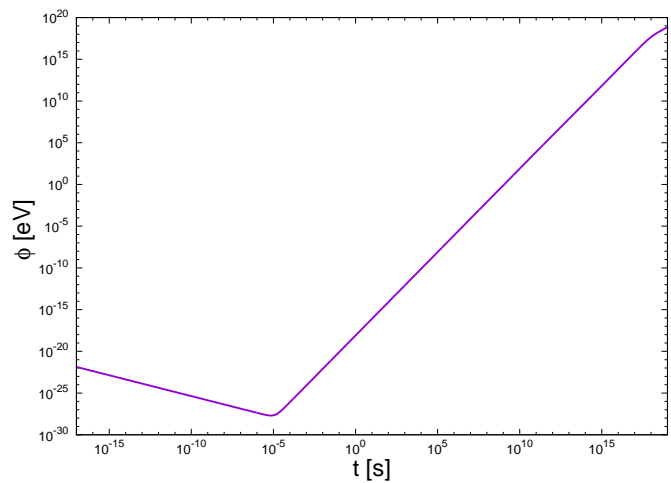}}
\caption{The development of $\varphi$. 
}\label{varphi}
\end{figure}

\begin{figure}[h]
\centering
\rotatebox{-90}{\includegraphics[width=15pc]{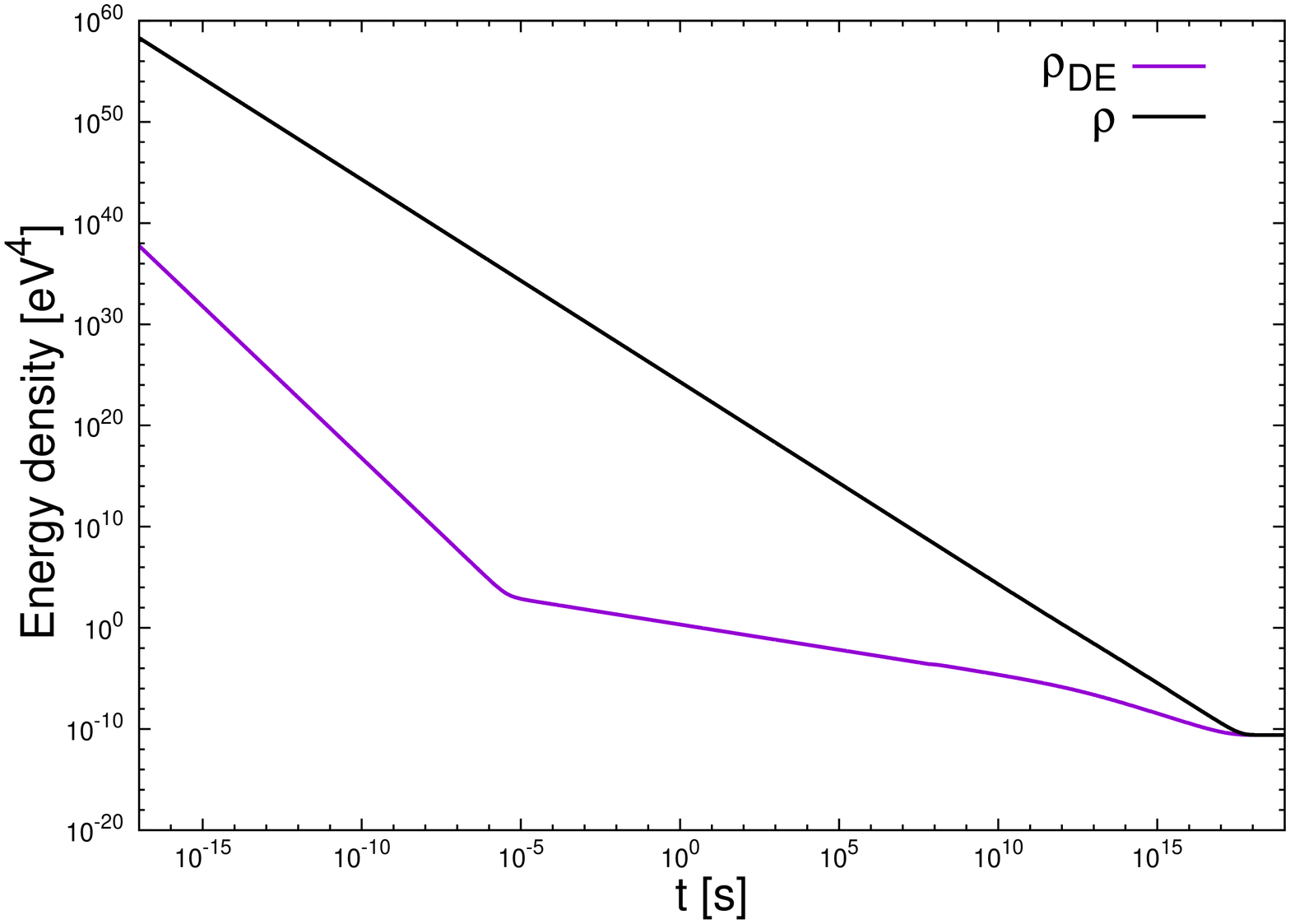}}
\caption{The development of energydensity. 
}\label{energy}
\end{figure}

We now solve the equations (\ref{SCCP3}), (\ref{SCCP4}), and (\ref{SCCP5}) 
numerically. In Fig.~\ref{lambda}, the time-development of $\lambda$ is given. 
The obtained value of $\lambda$ at the beginning of the radiation-dominated 
era is consistent with the analytic result in (\ref{lambdaC2}). 
In Fig.~\ref{varphi}, the time development of $\phi \equiv M_\mathrm{Pl}^3 \varphi$ is 
given. 
Fig.~\ref{energy} shows the development of the energy density. The parameters 
$\lambda_{\mathrm{rad}\, 1}$ and $\lambda_{\mathrm{rad}\, 2}$ are chosen to reproduce 
the value of the dark energy density in the current universe. 
The dark energy density in the matter-dominated era or the radiation-dominated era is 
surely negligible.

In summary, we have clarified the structure of the model in \cite{Nojiri:2016mlb} and 
investigated the cosmology given by the model. 
Although the model has an infinite numbers of the BRS symmetries, most of the symmetry 
is broken and there remains one and only one BRS symmetry which guarantee the unitarity 
of the model. We have also shown that by using the residual gauge symmetry,  the initial 
condition where $\varphi$ is a constant can be chosen. 
Because the quantum problem of the vaccum energy reduces to the classical problem of 
the initial condition in the model, we have investigated the region 
of the initial conditions which could be consistent with the evolution of the universe. 
It seems difficult to solve the fine-tuning problem in the initial condition in this model. 
It has been also shown that a stable solution describing the de Sitter space-time exist in this 
model.

\section*{Acknowledgments.}

The authors are indebted S. Akagi, K. Ichiki, T, Katsuragawa, R. Saitou and
N. Sugiyama. 
This work is supported (in part) by 
MEXT KAKENHI Grant-in-Aid for Scientific Research on Innovative Areas ``Cosmic
Acceleration''  (No. 15H05890) (D.N and S.N.).

\end{document}